\documentclass[letterpaper]{article}
\usepackage{tabularx} 
\usepackage{amsmath}  
\usepackage[font=small,labelfont=bf]{caption}
\usepackage{multicol}
\usepackage{longtable}
\numberwithin{equation}{section}
\usepackage{graphicx} 
\usepackage[margin=1in,letterpaper]{geometry} 
\usepackage{cite} 
\usepackage[affil-it]{authblk}
\usepackage{float}
\usepackage[final]{hyperref} 
\hypersetup{
	colorlinks=true,       
	linkcolor=blue,        
	citecolor=blue,        
	filecolor=magenta,     
	urlcolor=blue         
}

\makeatletter
\def\@maketitle{%
  \newpage
  \null
  \vskip 2em%
  \begin{center}%
  \let \footnote \thanks
    {\Large\bfseries \@title \par}%
    \vskip 1.5em%
    {\normalsize
      \lineskip .5em%
      \begin{tabular}[t]{c}%
        \@author
      \end{tabular}\par}%
    \vskip 1em%
    {\normalsize \@date}%
  \end{center}%
  \par
  \vskip 1.5em}
\makeatother

\begin{document}

\title{Comparisons of Matrices with Different  Elements  but Identical Eigenvalues}

\author{Daren Sitchepping Fosso}
\affil{School of Engineering, Rutgers University, Piscataway, NJ 08854}

\author{Larry Zamick}
\affil{Department of Physics and Astronomy, Rutgers University, Piscataway, NJ 08854}

\author{Castaly Fan}
\affil{Department of Physics, University of Florida, Gainesville, FL 32608}

\maketitle

\begin{abstract}
    
    We show 2 matrices that have identical eigenvalues but different eigenfunctions. This shows that in obtaining two body nuclear matrix elements empirically, it is not sufficient to consider only energy levels. Other quantities like transitions must also be included.

\end{abstract}

\section{Introduction}

    The nucleon-nucleon interaction is very complicated. In performing shell model calculations sometimes shortcuts are taken. For example one can take two body matrix elements from experiment as was done by McCullen, Bayman and Zamick \cite{PhysRev.134.B515}. See also Escuderos, Zamick and Bayman \cite{arxiv0506050} where improved matrix elements were used. They associated 2 body matrix elements in the $f_{7/2}$ shell $\langle f_{7/2}\, f_{7/2} |V| f_{7/2}\, f_{7/2} \rangle$ with excitation energies in the ``2 particle system'' $^{42}$Ca, $^{42}$S and $^{4}$Ti . When more configurations are included, the number of 2 body matrix elements grow by leaps and bounds.

    In this work we wonder if there is a danger in this procedure. We ask ``Can we have Hamiltonian matrices with identical eigenvalues (i.e. energy levels) but different eigenfunctions?'' With different eigenfunctions we would have different observable properties such as static electromagnetic moments, transition rates, half lives, etc. To answer this question we turn to simple matrices that were previously studied \cite{2018IJMPE..2750064Z, 2018IJMPE..2750087K, doi:10.1142/S021830131950037X, 2019IJMPE..2850062W, 2020IJMPE..2950050Z, 2021IJMPE..3050059F}. Mainly we consider tridiagonal matrices with one parameter [$v$], but also briefly consider pentadiagonal matrices with 2 parameters [$v,w$]. These will be shown in the next section.

\section{Results}

    We here show 2 pentadiagoal matrices $M_{1}$ and $M_{2}$:
    
    \begin{equation}
        M_{1} = 
        \begin{pmatrix}
        E_{0} & v & w & 0 & 0 & 0 & 0 & 0 & 0 & 0 & 0 \\
        v & E_{1} & v & w & 0 & 0 & 0 & 0 & 0 & 0 & 0 \\
        w & v & E_{2} & v & w & 0 & 0 & 0 & 0 & 0 & 0 \\
        0 & w & v & E_{3} & v & w & 0 & 0 & 0 & 0 & 0 \\
        0 & 0 & w & v & E_{4} & v & w & 0 & 0 & 0 & 0 \\
        0 & 0 & 0 & w & v & E_{5} & v & w & 0 & 0 & 0 \\
        0 & 0 & 0 & 0 & w & v & E_{6} & v & w & 0 & 0 \\
        0 & 0 & 0 & 0 & 0 & w & v & E_{7} & v & w & 0 \\
        0 & 0 & 0 & 0 & 0 & 0 & w & v & E_{8} & v & w \\
        0 & 0 & 0 & 0 & 0 & 0 & 0 & w & v & E_{9} & v \\
        0 & 0 & 0 & 0 & 0 & 0 & 0 & 0 & w & v & E_{10} \\
        \end{pmatrix}.
    \label{matrix1}
    \end{equation}

    \begin{equation}
        M_{2} = 
        \begin{pmatrix}
        E_{0} & -v & w & 0 & 0 & 0 & 0 & 0 & 0 & 0 & 0 \\
        -v & E_{1} & v & w & 0 & 0 & 0 & 0 & 0 & 0 & 0 \\
        w & v & E_{2} & -v & w & 0 & 0 & 0 & 0 & 0 & 0 \\
        0 & w & -v & E_{3} & v & w & 0 & 0 & 0 & 0 & 0 \\
        0 & 0 & w & v & E_{4} & -v & w & 0 & 0 & 0 & 0 \\
        0 & 0 & 0 & w & -v & E_{5} & v & w & 0 & 0 & 0 \\
        0 & 0 & 0 & 0 & w & v & E_{6} & -v & w & 0 & 0 \\
        0 & 0 & 0 & 0 & 0 & w & -v & E_{7} & v & w & 0 \\
        0 & 0 & 0 & 0 & 0 & 0 & w & v & E_{8} & -v & w \\
        0 & 0 & 0 & 0 & 0 & 0 & 0 & w & -v & E_{9} & v \\
        0 & 0 & 0 & 0 & 0 & 0 & 0 & 0 & w & v & E_{10} \\
        \end{pmatrix}.
    \label{matrix2}
    \end{equation}
    
    Both have on the diagonal $E(n) = n$. In matrix $M_{1}$ all the off-diagonal matrices are the same namely [$v,w$]. All other matrix elements are zero. We will first run the Mathematica program for the case $v = 1$ $w = 0$. This makes it a tridiagonal matrix. See Table \ref{tab:samev}.
    
    In matrix 2 $M_{2}$ the first off diagonal matrix elements alter in sign $(-1)^{n+1}v$ and we still have $w=0$. All other matrix elements are zero. We will later consider cases where both and $w$ are non-zero. See Table \ref{tab:differentv}.

    Both $M_{1}$ and $M_{2}$ are 11 by 11 matrices. The matrix $M_{1}$ was previously studied \cite{2018IJMPE..2750064Z, 2018IJMPE..2750087K, doi:10.1142/S021830131950037X, 2019IJMPE..2850062W, 2020IJMPE..2950050Z, 2021IJMPE..3050059F}. We here first run the case $v = 1$, $w=0$ but other values are easy to obtain as well. We find that the eigenvalues of the 2 matrices, $M_{1}$ and $M_{2}$, are identical.

    Let the eigenfunctions of $M_{1}$ be $a(n,m)$ the $m$'th component of the $n$'th eigenfunction of $M_{1}$ and $b(n,m)$ of $M_{2}$. Then we find that the magnitudes of the $b(n,m)$ is the same as that of $a(n,m)$ but the phases are different.

    Here we show a few eigenvalues and eigenfunction components which will illustrate the similarities and differences.

    \begin{table}[H]
    \centering
    \caption{The numerical result of $M_{1}$.}
    {\renewcommand{\arraystretch}{1.25}
    \begin{tabular}{c|c|c c c c}
            state & eigenvalue & $a_{0}$ & $a_{1}$ & $a_{2}$ & $a_{3}$\\
            \hline
            0 & -7.46E-1 & 7.77E-1 & -5.79E-1 & 2.35E-1 & -6.67E-2\\
            1 & 7.89E-1 & -5.41E-1 & -4.27E-1 & 6.31E-1 & -3.37E-1\\
            2 & 1.96 & 2.99E-1 & 5.85E-1 & 2.64E-1 & -5.96E-1\\
            3 & 2.99 & -1.16E-1 & -3.48E-1 & -5.79E-1 & -2.28E-1\\
        \end{tabular}}
    \label{tab:samev}
    \end{table}

    \begin{table}[H]
    \centering
    \caption{The numerical result of $M_{2}$.}
    {\renewcommand{\arraystretch}{1.25}
    \begin{tabular}{c|c|c c c c}
            state & eigenvalue & $a_{0}$ & $a_{1}$ & $a_{2}$ & $a_{3}$\\
            \hline
            0 & -7.46E-1 & 7.77E-1 & 5.79E-1 & -2.35E-1 & -6.67E-2\\
            1 & 7.89E-1 & 5.41E-1 & -4.27E-1 & 6.31E-1 & 3.37E-1\\
            2 & 1.96 & -2.99E-1 & 5.85E-1 & 2.64E-1 & 5.96E-1\\
            3 & 2.99 & -1.16E-1 & 3.48E-1 & 5.79E-1 & -2.28E-1\\
        \end{tabular}}
    \label{tab:differentv}
    \end{table}

    We next list for the first 4 states whether the signs for the 2 cases (same $v$ and alternating $v$) are the same (S) or are different (D), as shown in Table \ref{signv}.

    \begin{table}[H]
        \centering        
        \caption{Relative signs for the cases same $v$ and alternating sign $v$.}
        \begin{tabular}{l|l l l l}
            state &  signs\\
            \hline
            0 & $SDD$ & $SSD$ & $DSS$ & $DD$\\
            1 & $DSS$ & $DDS$ & $SDD$ & $SS$\\
            2 & $DSS$ & $DDS$ & $SDD$ & $SS$\\
            3 & $SDD$ & $SSD$ & $DSS$ & $DD$\\
            4 & $DSS$ & $DDS$ & $SDD$ & $SS$\\
            5 & $DSS$ & $DDS$ & $SDD$ & $SS$\\
            6 & $SDD$ & $SSD$ & $DSS$ & $DD$\\
            7 & $DSS$ & $DDS$ & $SDD$ & $SS$\\
            8 & $DSS$ & $DDS$ & $SDD$ & $SS$\\
            9 & $DSS$ & $DDS$ & $SDD$ & $SS$\\
            10 & $DSS$ & $DDS$ & $SDD$ & $SS$\\
        \end{tabular}
        \label{signv}
    \end{table}

    There appear to be 2 sets of phase differences. For eight of them the pattern is $DSS$ $DDS$ $SDD$ $SS$, whist for three it is $SDD$ $SSD$ $DSS$ $DD$. However it should be remembered that if we multiply a wave function by a minus sign we get the same wave function. Hence we can say that all patterns are the same.
    
    We can explain the identical eigenvalues by considering the smallest tridiagonal 3 by 3 matrix:
        \begin{equation*}
            \begin{pmatrix}
                E_{0} & -v & 0\\
                -v & E_{1} & v\\
                0 & v & E_{2}
            \end{pmatrix}
        \end{equation*}

    Let the eigenvalues be $\lambda$. Define $B_{0} = E_{0} - \lambda$; $B_{1} = E_{1} - \lambda$; $B_{2} = E_{2} - \lambda$.
    To get the eigenvalues we set the determinant of the following matrix to zero.
        \begin{equation*}
        \begin{pmatrix}
            B_{0} & -v & 0\\
            -v & B_{1} & v\\
            0 & v & B_{2}
        \end{pmatrix}
        \end{equation*}
    We obtain the following equation:
    \begin{equation}
        B_{0} B_{1} B_{2} - B_{0} v^{2} -B_{2} v^{2} =0
    \end{equation}
    
    We don't have to solve this . Just note that even though our matrix has both $+v$'s and $-v$'s the above equation has only $v^{2}$ in the expression. Clearly we will get the same quadratic equation for the matrix with all $+v$'s. It is also easy to see that the eigenfunctions will be different in the 2 cases.

\section{Lifetime Differences}

    These 11 by 11 matrices above, $M_{1}$ and $M_{2}$, can be regarded as simulated quantum mechanical Hamiltonians. The eigenvalues are associated with discreet energy levels. We add to this transition rates. We know that in an a atom or a nucleus if the system is in state $|N_{2}\rangle$ it can go to a lower state $|N_{1}\rangle$ by emitting a photon (i.e. a quantum of light) of energy ($E_{2}-E_{1}$).

    The frequency (i.e. color) of the photon is $f=(E_{2}-E_{1})/h$. Where $h$ is Planck's constant. We had in previous works defined for our simulated Hamiltonian a simulated transition operator $O$. Such that for
        \begin{equation}
            O = a_{0} b_{1} + a_{1} b_{2} + a_{2} b_{3}+ ... +a_{9} b_{10} + a_{1} b_{0} + a_{2} b_{1} + a_{3} b_{2} +. ..+a_{10} b_{9}
        \end{equation}
    In the above the $a_{n}$'s are the eigenfunction components of one state, the $b_{n}$'s are the eigenfunction components of the other state, and the transition is from state $|a\rangle$ to state $|b\rangle$.

    We then note that the transition rate is proportional to $|O|^2$ and the half-life is proportional to $1/|O|^2$. 
    
    We now return to the point that for the 2 matrices, which we now call simulated Hamiltonians. The lifetimes will be different for the cases where the off diagonals are $(v,v,v,v,v,v,v...)$ or are $(v,-v,v,-v,...)$. This is despite the fact that the eigenvalues are identical.

    We show results in Table \ref{lifetime} for the ratio of lifetimes for alternating sign $v$'s versus to same sign. We assume the dominant transition is from $|n+1\rangle$ to $|n\rangle$.

    \begin{table}[H]
        \centering        
        \caption{Ratio of lifetimes.}
        \begin{tabular}{c|c}
            energy levels & ratio of lifetimes\\
            \hline
            From 1 to 0 & 3.34\\
            From 2 to 1 & 0.75\\
            From 3 to 2 & 1.50\\
            From 4 to 3 & 1.69\\
            From 5 to 4 & 1.73\\
            From 6 to 5 & 1.73\\
            From 7 to 6 & 1.69\\
            From 8 to 7 & 1.50\\
            From 9 to 8 & 0.75\\
            From 10 to 9 & 3.34\\
        \end{tabular}
        \label{lifetime}
    \end{table}

\section{Other Comparisons for a Tridiagonal Matrices}

    We can make a simpler transformation for $M_{1}$. We consider 0 to $n$ transitions. We compare [$v=1; w=0$] with [$v=-1; w=0$].

    The eigenvalues do not change when we change the sign of $v$. We find that the magnitude of $\langle n | O |m\rangle$ for $n$ not equal to m does not change. In some cases the sign changes and sometimes not. But this makes no difference for what one can measure -- a transition rate or half-life. These involve the square of $\langle n | O |m\rangle$.

    For the case $\langle n | O |n\rangle$ we find that there is a change in sign ask we change $v$ from $+1$ to $-1$. This leads to measurable consequences -- namely static moments. We have a situation here where energy levels and transition rates are not affected by changing the sign of $v$. The only thing that is sensitive is the static moment.

\section{Interesting Results for Simple Pentadiagonal Matrices}

    In this section we consider very simple pentadiagonal matrices. Referring  to $M_{1}$ of Eq. \ref{matrix1} we consider  and compare two cases  [$v=0, w=1$]and [$v=0,w=-1$].
    
    Let us state at the outset that the eigenvalues for the cases are identical. In Table \ref{tab:transition} we show the results for matrices of various sizes $4\times 4$, $6\times 6$, $8\times 8$, and $10\times 10$. The transition amplitudes from $n=0$ to even $n$ are all zero so we need only consider transitions to odd $n$. 

     \begin{table}[H]
    \centering
    \caption{Transition amplitudes from $n=0$.}
    Four by four, from $n=0$\\
    {\renewcommand{\arraystretch}{1.25}
    \begin{tabular}{p{1cm}|p{3cm} p{3cm}}
            $n$ & $t=+1$ & $t=-1$\\
            \hline
            1 & -6.464E-1 & 1.354\\
            3 & -1.464E-1 & -1.464E-1\\
        \end{tabular}}
    \vspace{3em}
    
    Six by six, from $n=0$\\
    {\renewcommand{\arraystretch}{1.25}
    \begin{tabular}{p{1cm}|p{3cm} p{3cm}}
            $n$ & $t=+1$ & $t=-1$\\
            \hline
            1 & 5.918E-1 & 1.488\\
            3 & 9.175E-2 & -9.175E-2\\
            5 & 0 & 0\\
        \end{tabular}}
    \vspace{3em}
    
    Eight by eight, from $n=0$\\
    {\renewcommand{\arraystretch}{1.25}
    \begin{tabular}{p{1cm}|p{3cm} p{3cm}}
            $n$ & $t=+1$ & $t=-1$\\
            \hline
            1 & -5.884E-1 & 1.412\\
            3 & -8.390E-2 & -8.390E-2\\
            5 & -1.431E-2 & -1.431E-2\\
            7 & 3.010E-3 & 3.010E-3
        \end{tabular}}
    \vspace{3em}
    
    Ten by ten, from $n=0$\\
    {\renewcommand{\arraystretch}{1.25}
    \begin{tabular}{p{1cm}|p{3cm} p{3cm}}
            $n$ & $t=+1$ & $t=-1$\\
            \hline
            1 & 5.883E-1 & 1.417\\
            3 & 8.351E-2 & 8.351E-2\\
            5 & 1.541E-2 & 1.541E-2\\
            7 & 2.098E-3 & -2.098E-3\\
            9 & 0 & 0
        \end{tabular}}
    \label{tab:transition}
    \end{table}

    To gain insight into what is happening we will consider a reduced 6 by 6 matrix as shown.
    \begin{equation}
        M_{6} = 
        \begin{pmatrix}
            E_{0} & v & w & 0 & 0 & 0\\
            v & E_{1} & v & w & 0 & 0\\
            w & v & E_{2} & v & w & 0\\
            0 & w & v & E_{3} & v & w\\
            0 & 0 & v & w & E_{4} & v\\
            0 & 0 & 0 & v & w & E_{5}
        \end{pmatrix}
    \end{equation}
    Here $E_{n} =n$, $v=0$ and in one case we have  $w=+1$ and in the other $w= -1$. There is no loss in generality by considering this smaller matrix.

    The eigenfunctions in $M_{6}$ are of the form
    \begin{equation*}
        d_{0} |0\rangle +d_{1} |1\rangle + d_{2} |2\rangle + d_{3} |3\rangle + d_{4} |4\rangle +d_{5} |5\rangle + d_{6} |6\rangle.
    \end{equation*}
    We also defined a transition amplitude from say, $|D\rangle$ to $|G \rangle$
    \begin{equation}
        O(D,G)= d_{0} g_{1} +d_{1} g_{2} +d_{2} g_{3} +d_{3} g_{4} +d_{4} g_{5} +d_{5} g_{6} +g_{0} d_{1}+g_{1} d_{2} +g_{2} d_{3} +g_{3} d_{4} + g_{4} d_{5} +g_{5} d_{6}.
    \end{equation}
    We compare results for eigenvalues, eigenfunctions and transition amplitudes for the 2 cases [$v=0, w=1$] and [$v=0, w=-1$]. 
    
    Here are the eigenfunctions components $|n\rangle$ for the case [$v=0, w=1$]. We only show them for odd $n$ because it can be shown that transition amplitudes from $n=0$ to even $n$ are zero.

    Define $D(n) = |d(n)|$. For $[v=0,w=+1]$,
    \begin{equation*}
    \begin{split}
        D_{0} &= 0.908248\\
        D_{1} &= 0.408248\\
        D_{2} &= 0.091752\\
        D_{3} &= 0.816497
    \end{split}
    \end{equation*}
    The wave function components are
    \begin{equation*}
    \begin{split}
        n&=0: D_{0}, 0 , -D_{1}, 0, D_{2}, 0\\
        n&=1: 0 , D_{0}, 0, -D_{1}, 0, D_{2}\\
        n&=3: 0 , -D_{1}, 0, -D_{3} , 0, D_{1}\\
        n&=5: 0 , D_{2}, 0, D_{1} , 0, D_{0}
    \end{split}
    \end{equation*}
    The transition amplitudes for $w+1$ are
    \begin{equation}
    \begin{split}
        O( 0 \rightarrow 1) &= 1-D_{0} D_{1} -D_{1} D_{2}\\
        O( 1 \rightarrow 3) &= D_{1}D_{1} -D_{3} D_{2} - D_{0} D_{1} +D_{1} D_{3} +D_{2} D_{1}\\
        O( 1 \rightarrow 5) &= -D_{1}D_{1} +2 D_{0} D_{2}
    \end{split}
    \end{equation}
    
    There are a couple of interesting results that are immediately apparent.
    
    From $n= 0$ to $n=1$ in the transition amplitude for $[0,1]$ has magnitude $1-X$ while for $[0,-1]$ it is $1+X$ (the sum is two). The 2 transition amplitudes are quite different, even though the eigenvalues are the same.
    
    From $n= 0$ to $n=3$ the two transition amplitudes look quite different. However we claim that $-D_{0} D_{1} + D_{1} D_{3} +D_{2} D_{1} =0$. This stems form the fact that the $n=1$ state is orthogonal to the $n=3$ state. We now see that the two $n=3$ transition amplitudes are equal and opposite, with the $[0,+1]$ expression being $D_{1} D_{1} - D_{3} D_{2}$. 
    
    We can now guess at a generalization to a matrix of any size that the only case in which the magnitudes of the transitions amplitudes are different is the transition from $n= 0$ to $n =1$. In all other cases the magnitudes are identical, which is a quite fascinating behavior.
    
    From $n= 0$ to $n=5$ we see that the transition amplitudes are also the same. In fact both are zero. For $n= 5$ the transition amplitude is $O( 1 \rightarrow 5) = -D_{1}D_{1} +2 D_{0} D_{2}$. But this is also the expression for the overlap of the $n=0 $and $n=5$ state $\langle n=0|n=5 \rangle$. But this is clearly zero—the orthogonality relation for different eigenstates. It turns out that one gets. zero transition from $n=0$ to $n_{\text{max}}$ for $n_{\text{max}} =2,6,10$, e.t.c.

    In all of the cases in Table \ref{tab:transition}, $4\times 4$ to $10\times 10$, we find that the magnitudes of the transition amplitudes for the 2 cases [0,1] and [0,-1] from $n=0$ to $n=3$ or higher are identical. However from $n=0$ to $n=1$ they are quite different. Indeed we have the remarkable results in this case that the sum of the magnitudes for [0, 1] and [0,-1] is two.

\section{More Complex Pentadiagonal Matrices}

    We again focus on $M_{1}$ pentadiagonal -- zero to $n$ transitions. We consider 2 sets:
    \begin{enumerate}
        \item Set 1: [$v=1; w=1$], [$v=-1; w=1$]
        \item Set 2: [$v=1; w=-1$], [$v=-1; w=-1$]
    \end{enumerate}
    We find that the eigenvalues for the 2 members of Set 1 have identical eigenvalues.
    The eigenvalues of the 2 members of Set 2 are also identical but they are different from those of Set 1.

    The magnitudes of the transition matrix elements $\langle 0 | O |n\rangle$ for the 2 members of Set 1 are the same.The phases are sometimes the same and sometimes different. But since the the transition rates and half lives involve the square of $\langle 0 | O |n\rangle$ they will be the same for the 2 members. A similar story for Set 2 the magnitudes of $\langle 0 | O |n\rangle$ are the same for the 2 members. And so the transition rates and half lives are the same. But the transition rates for Set 1 are different that those of Set 2.

\section{Using the Shell Model to get Matrices with Identical Eigenvalues}

    Consider 2 neutrons that can be either in the $f_{7/2}$ or $p_{3/2}$ shell. Consider a state with total angular momentum $J=2$. It can be written as a linear combination of 3 configurations
    
    \begin{equation}
    \begin{split}
        |1\rangle &= [f_{7/2} f_{7/2}] J=2\\
        |2\rangle &= [p_{3/2} p_{3/2}]J=2\\
        |3\rangle &= \frac{1}{\sqrt{2}}(1-P_{12})[f_{7/2}p_{3/2}]J = 2
    \end{split}
    \end{equation}
    where $P_{12}$ is the particle exchange operator. We get a secular matrix $H_{mn} = \langle m |V| n\rangle$. In the single particle configuration you can choose the radial wave function to be either positive near the origin. If we change from positive to negative for $p_{3/2}$ then $|3\rangle$ will change sign but $|1\rangle$ and $|2\rangle$ will not.

    So we have 2 matrices
    \begin{equation}
        M'_{1} = \begin{pmatrix}
            H_{11} & H_{12} & H_{13}\\
            H_{21} & H_{22} & H_{23}\\
            H_{31} & H_{32} & H_{33}
        \end{pmatrix};
    \end{equation}
    and
    \begin{equation}
        M'_{2} = \begin{pmatrix}
            H_{11} & H_{12} & -H_{13}\\
            H_{21} & H_{22} & -H_{23}\\
            -H_{31} & -H_{32} & H_{33}
        \end{pmatrix}.
    \end{equation}
    Since changing the overall sign of a given single particle state should make no difference, we expect the eigenvalues to be the same for the 2 matrices.
    
    As an example consider the following:
    
    \begin{equation*}
        M'_{1} = \begin{pmatrix}
            0 & 3 & 1\\
            3 & 1 & 4\\
            1 & 4 & 2
        \end{pmatrix}, \;
        M'_{2} = \begin{pmatrix}
            0 & 3 & -1\\
            3 & 1 & -4\\
            -1 & -4 & 2
        \end{pmatrix}.
    \end{equation*}

    The three eigenvalues are: $-3.48134$, $-0.214496$ and $6.69584$ for both $M'_{1}$ and $M'_{2}$. The eigenfunctions are however different. One can take two views from here.

    For $M'_{1}$ they are:
    \begin{equation*}
    \begin{split}
        &0.5071 |1\rangle -0.7375 |2\rangle + 0.4456 |3\rangle\\
        &0.7679 |1\rangle + 0.1525 |2\rangle -0.6222 |3\rangle\\
        &0.3909 |1\rangle +0.6579 |2 \rangle +0.6437 |3 \rangle
    \end{split}
    \end{equation*}
    
    For $M'_{2}$ they are :
    \begin{equation*}
    \begin{split}
        &0.5071 |1\rangle -0.7375 |2\rangle -0.4456 |3\rangle\\
        &0.7679 |1\rangle + 0.1525 |2\rangle +0.6222 |2\rangle\\
        &-0.3909 |1\rangle -0.6579 |2\rangle +0.0.6437 |3\rangle
    \end{split}
    \end{equation*}
    
    We can now take two different points of view. First there was no change in the radial wave function $|3\rangle$. There are genuinely different matrix elements competing to explain properties the system e.g. transition rates. In that case the two matrices will yield different results. The second point of view is that the difference is due to having changed the radial wave function. In that case on has to change the signs of the one body transition operators accordingly. In that case the transition rates for the 2 matrices will be the same.

\section{Closing Remarks}

    We have successfully found matrices with identical eigenvalues and different eigenfunctions. In some cases e.g. when all we do is change $v$ from $+1$ to $-1$ in $M_{1}$ we get different phases for transition matrices. However since transition rates involve squares of transition matrix elements, one will not see changes in these rates.

    Static properties will undergo changes of sign e.g. magnetic moments and quadrupole moments. However for the more elaborate changes, e.g. going from $M_{1}$ with all $v$'s the same, to $M_{2}$ with alternating $v$'s we do get differences in transition rates and hence half-lives.

    In our brief encounter with pentadiagonal matrices we also find cases of matrices with identical eigenvalues and different eigenfunctions. This suggest that there there are more complex matrices than what we have so far considered with the dual properties -- same eigenvalues but different eigenfunctions. We hope this work will stimulate further work on this subject.

    Our work shows the danger of obtaining nuclear matrix elements by only fitting energy levels. The differing wave functions shown lead to other different properties e.g. different transition rates of decay for excited levels. So those quantities should also be included when fits to nuclear interaction matrix elements are made.

    Daren Sitchepping Fosso received support from the Louis Stokes Alliance for Minority Participation (LSAMP).

\nocite{*} 
\bibliographystyle{unsrt}
\bibliography{main.bib}

\end{document}